\documentclass[10pt,twocolumn,letterpaper]{article}

\usepackage[T1]{fontenc}
\usepackage[utf8]{inputenc}
\usepackage{mathptmx}
\usepackage[letterpaper,top=0.75in,bottom=1in,left=0.625in,right=0.625in,
            columnsep=0.25in]{geometry}
\usepackage{graphicx}
\usepackage{array}
\usepackage{booktabs}
\usepackage{tabularx}
\usepackage{ragged2e}
\usepackage{caption}
\usepackage{titlesec}
\usepackage{xcolor}
\usepackage{tikz}
\usetikzlibrary{arrows.meta,positioning,fit,backgrounds}
\usepackage[hidelinks]{hyperref}
\usepackage{authblk}

\usepackage{microtype}

\renewcommand{\thesection}{\Roman{section}}
\renewcommand{\thesubsection}{\Alph{subsection}}

\titleformat{\section}[block]{\centering\normalsize\scshape}{\thesection.}{0.5em}{}
\titleformat{\subsection}[block]{\itshape\normalsize}{\thesubsection.}{0.5em}{}
\titlespacing{\section}{0pt}{1.2em}{0.6em}
\titlespacing{\subsection}{0pt}{0.9em}{0.4em}

\newcolumntype{L}[1]{>{\RaggedRight\arraybackslash}p{#1}}

\title{\vspace{-2em}\bfseries\large When Attention Guardrails Become Barriers
to Learning: Towards the Tipping Point}
\author[1,*]{Meenakshi~V}
\author[2]{Pavani~Ayinampudi}
\author[2]{Aditya~B.M.V.}
\author[2]{Jinal~Gupta}
\author[2]{Prakash~Hegade}
\author[1]{Rohit~Sharma}
\author[1]{Sakshi~Sharma}
\author[1]{S.R.S.~Iyengar}
\affil[1]{Indian Institute of Technology Ropar, Rupnagar, Punjab, India\\
  \texttt{\{meenakshi.19csz0013, rohit.24csz0014, sakshi.23csz0006,
  sudarshan\}@iitrpr.ac.in}}
\affil[2]{ANNAM.AI, Indian Institute of Technology Ropar, Rupnagar, Punjab, India\\
  \texttt{\{pavania.harvard2025, adityabmv, jinalbirla,
  prakash.hegade\}@gmail.com}}
\affil[*]{Corresponding author: \texttt{meenakshi.19csz0013@iitrpr.ac.in}}
\date{}

\begin{document}
\twocolumn[
  \begin{@twocolumnfalse}
    \maketitle
    \begin{abstract}
      \noindent\small
      Online learning offers flexibility but lacks the structure of a
      classroom, where a teacher's presence guides attention. The platform we
      study restores that structure by monitoring the learner through the
      webcam during ordinary coursework, interrupting or restarting a video
      when the learner appears distracted. What such monitoring does to a
      learner across a whole course, rather than in a single examination, is
      largely unexamined. We report a convergent mixed-methods study of one
      monitored course pipeline in a summer internship program. We read two
      free-text surveys alongside three platform channels. The dropout exit
      survey yielded 15 analyzable responses; the persisting-learner
      reflection survey, 36. The channels are camera-verification telemetry
      (14,529 flags from 448 students), an in-video emotion widget (615
      submissions from 273 students), and a mandatory end-of-course survey
      (up to 634 respondents per item). Neither survey named monitoring, so
      every mention analyzed here was raised by the respondent.
      Focus-monitoring was raised by 18 of the 51 free-text respondents: 7 of
      15 dropouts and 11 of 36 persisting learners. Among the dropouts who
      raised it, focus-monitoring was the stated primary cause of departure
      in 4 of 7 cases. None of the 18 questioned being observed in principle. What
      learners contest is the misreading of ordinary actions---drinking water
      or moving the head---and the severity of what follows a flag: a video
      already watched returns to the start of its segment. These findings identify two targets for redesign: the severity of the
      response to a flag, and the environment check, which can flag a learner
      before any content has been seen. We argue that the proportionality of
      that response marks the point at which an attention guardrail becomes a
      barrier to learning, the tipping point this study approaches.
      \vspace{0.6em}

      \vspace{1.2em}
    \end{abstract}
  \end{@twocolumnfalse}
]

\section{Introduction}

Education, in its earliest Indian form, was a lived apprenticeship. Under the
Guru-Shishya parampara, the learner resided with the guru, and learning
unfolded not in scheduled hours but through every shared moment of the day,
under the guru's watchful presence. The modern classroom is our first attempt
to scale that tradition. One teacher now meets many students, but within four
walls deliberately arranged for attention: a well-lit room, proper seating,
peers on every side, and a teacher who can see every face and sense, in
person, whether the material is being understood. The classroom is, above all, a
controlled environment, built so that a learner's attention has somewhere to
rest and few places to escape.

Online education has scaled that tradition even further, enabling learners to
enter the
classrooms of renowned professors from institutions that they can never reach
physically. However, the self-paced online course does not reproduce the
structured environment that the classroom had built for the control of
attention: the fixed schedule, the peers working alongside, the daily contact
with a teacher, and the room kept free of distraction. What remains is
content, delivered to a learner who must now provide all the structure alone.
Completion rates reflect this absence of structure. Most massive open online courses are
completed by fewer than 13\% of those who enroll (Onah et al., 2014). Studies
of why learners leave repeatedly point to motivation, time management, and the
absence of interaction rather than the content itself (Wang et al., 2023;
Azhar et al., 2024).

Meanwhile, the environment in which online learning happens has grown more
hostile to attention, not less. A student watching a lecture at home watches
it on the same device that carries every notification, and multitasking is
measurably higher in online courses than in face-to-face classrooms (Lepp et
al., 2019; Deng et al., 2024). In a classroom, a teacher notices a wandering
gaze and draws it back. At home, no one performs that function. Hence the
question every online platform must now answer: content is being delivered,
but is a structured learning environment also being provided?

One emerging answer is to let the learner's own webcam do what the teacher's
eye once did. A growing body of work builds systems that detect attention and
engagement from webcam video during online learning (Dewan et al., 2019; Robal
et al., 2018a), and such monitoring is beginning to move out of the laboratory
and into learning platforms themselves. We use the term
\emph{focus-monitoring} for this class of mechanisms. The platform observes
the learner through camera and microphone not to proctor an examination, but to
act as an attention guardrail, noticing distraction and guiding the learner back, the way
a teacher or a parent once would. A course is \emph{verified} in the sense we
use throughout this paper when continued playback depends on the platform
confirming, moment to moment, that the enrolled learner is present and
attending. This is the motivation behind ViBe, an AI-based learning platform that embeds focus-monitoring into
everyday course delivery.

However, building the guardrail is only half the problem, and it is the half
the literature has concentrated on. The technical work asks whether attention
can be detected. The proctoring literature asks whether monitoring imposes a
cost on test-takers (Conijn et al., 2022; Marano et al., 2024). What remains
inadequately understood is the learner's side of the experience in an ordinary
course, outside the examination hall. Existing work does not examine what
learners require to remain focused and complete an online course, nor how they
respond when a platform provides such support. To address these gaps, we study
the experience of students who took courses on ViBe and lived with its focus-monitoring throughout, drawing on
accounts from both the learners who persisted through the coursework and the
learners who left it. This paper examines what focus-monitoring does to a learner's
experience of a course, and which of its mechanisms carry the cost. Our
interest is the tipping point: the moment at which an attention guardrail,
built to hold a learner in a course, becomes a barrier to learning it.

The rest of the paper is organized as follows. Section II reviews related
work. Section III describes the methodology. Section IV presents the findings.
Section V discusses their implications. Section VI states the limitations, and
Section VII concludes and outlines the future work.

\section{Related Work}

Success in a self-paced online course rests on how well the learner can
regulate their own learning. Self-regulated learning (SRL) describes this
capacity, the ability to plan, monitor, and control one's own cognition,
motivation, and behavior in pursuit of learning goals (Zimmerman, 2000), with
motivation sustaining that regulation over time (Pintrich, 1999). Kizilcec et
al. (2017), studying 4,831 learners across six MOOCs, find that SRL skills
predict whether learners attain their course goals, because MOOCs offer low
support and leave regulation entirely to the learner, though their study
measures the skills learners bring rather than testing whether a platform
could provide regulation to those who lack it. Reparaz et al. (2020) reach the
complementary finding from the retention side: completers self-regulate
significantly better than non-completers, but the design is correlational and
offers the struggling learner nothing beyond the observation. This framework
anchors our study: a platform that provides regulation externally acts as a
scaffold for SRL, and whether learners experience it that way is the question
we take up.

The regulation these courses demand has meanwhile become harder to sustain.
Digital devices are now the primary source of distraction in learning
contexts, and students' ability to resist them varies with motivation and
self-regulation skill (Deng et al., 2024), though that study examines the
physical classroom, where a teacher's presence still moderates behavior. In a
self-paced online course, no teacher is present to provide this moderation. Lepp et al. (2019) find college
students multitask more in online courses than in face-to-face settings,
though their design includes no platform intervention. Similarly, Wu (2017)
links media multitasking to degraded performance, though this relies on
learner perception rather than observed behavior. Across these studies,
distraction emerges as a structural condition of online learning rather than
an individual failing, yet in every case the learner is left to manage their attention. 
The platform under study answers this structurally: attention monitoring is
embedded within the learning environment itself rather than offered as an
occasional or optional intervention.

Attention monitoring entered the education system through the examination hall. Conijn et
al. (2022), surveying 1,760 students across 105 courses, found that proctored
examination raises test anxiety while leaving performance unaffected, but the
study measures the exam sitting as a whole and cannot say which behavior of
the monitoring system produces the cost. Woldeab and Brothen (2019) narrow
the anxiety finding to students with trait test anxiety, again in the
examination setting only. In both, monitoring is an event of a few high-stakes
hours, not a condition of everyday learning.

More recently, monitoring has moved from the examination hall into learning
itself. Dewan et al. (2019) review engagement-detection methods and conclude
that webcam-based approaches are promising and non-intrusive, though the
systems reviewed remain largely research prototypes. Robal et al. (2018a) test
the feasibility of webcam-based attention-loss detection, motivated explicitly
by MOOC learners' weak self-regulation, but evaluate tracking frameworks
against a benchmark rather than in a running course. The same team then built
IntelliEye (Robal et al., 2018b), the closest prior system to ViBe: real-time attention tracking during MOOC lecture
videos, deployed in a live course for 74 days, alerting distracted learners
through visual and auditory cues. The deployment measured acceptance and
behavior change, but participation was voluntary and the study does not
connect monitoring to completion. Hutt et al. (2021) carry the intervention
furthest, with gaze-based mind-wandering detection driving real-time
re-engagement prompts for 287 high-school students, improving attention and
retention for some learners, but in a physical classroom with dedicated
eye-tracking hardware and a teacher present, the opposite of the unsupervised,
self-paced setting we examine.

Existing research largely focuses on examination monitoring or optional aids
for persisting learners. Marano et al. (2024) identify a gap regarding
learners who disengage partway, and no prior work examines mandatory,
course-long focus-monitoring. We
address this gap by studying both the learners who persisted and those who
left, in their own words. The next section presents the methodology.

\section{Methodology}

\subsection{Objective and Research Questions}

This study investigates how learners experience persistent focus-monitoring in
everyday coursework. While the literature confirms that such systems can be deployed,
their impact on learners across an entire course remains unexamined. One
overarching question guides the study, approached through two sub-questions:

\vspace{0.4em}
\noindent\textbf{RQ}: How does platform-enforced focus-monitoring influence
learners' engagement and persistence in self-paced online courses?

\noindent\textbf{RQ1}: Do learners perceive the monitoring as a support layer
that helps them stay focused and complete the course?

\noindent\textbf{RQ2}: In what ways, and to what extent, does the monitoring
hinder them?
\vspace{0.4em}

To answer these questions, we study one cohort of learners in a single
monitored course delivered on ViBe, a
platform in which focus-monitoring is embedded in everyday course delivery
rather than reserved for examinations. The learners divide into those who left
that course and those who completed it and continued into the next course.

\subsection{The Platform and the Course}

ViBe is an online learning platform built
around the principle of continuous active learning: a course is a fixed
sequence of short video segments and quizzes, and a learner who answers
incorrectly is sent back to rewatch the relevant video before moving on. The
design draws its name from the Indian tale of Vikram and Betaal, in which
every wrong answer returns the seeker to the beginning. Progression is
strictly linear. A learner cannot skip ahead to later content, and cannot seek
forward within a video past the furthest point already watched.

Focus-monitoring on ViBe begins before the
first video plays. A learner grants camera and microphone permission,
registers a reference photograph of their face, accepts a consent form and a
proctoring declaration, and enters the course in mandatory fullscreen. During
playback, the webcam is checked continuously for the learner's
presence, for additional faces in the frame, for a match against the
registered face, and for a blurred or obstructed view, while the microphone is
monitored for speech. When the system detects an anomaly, a full-screen alert
covers the video, showing the learner their own webcam feed and the reason for
the interruption, and clears only when the anomaly does. A sustained anomaly
of roughly two seconds, or the detection of a second person or an unrecognized
face, pauses the video and rewinds it to the beginning of the current segment.
Each flagged moment also captures a webcam snapshot for instructor review.
Playback speed may be raised up to twice the normal rate, but no content can
be skipped unwatched. Quizzes allow unlimited attempts; a failed attempt
routes the learner back to the preceding video, and assessment scores below a
required threshold can force a restart of the content sequence.

The studied coursework is part of a summer internship program for Indian
undergraduates. Interns complete an onboarding course, then `Fundamentals of
AI using agriculture datasets' (approximately one week), and then `MERN stack
development'. The dropout accounts in this study concern learners who did not
continue the Fundamentals of AI course, the first content course in the
pipeline; the persisting-learner accounts come from learners who progressed
beyond it into the MERN course. We refer to this group as persisting learners
throughout, meaning learners who continued past the first content course,
whether or not they had completed the course that followed. Completion of a
course required finishing every item in its sequence, and the completion
certificate was issued by the program.

Throughout this paper, focus-monitoring refers to these mechanisms as one
bundle: camera-based presence and identity checking, microphone
monitoring, distraction-triggered rewinds, assessment gating, and sequencing
enforcement. The bundle is read as one construct because every mechanism in it
serves the same function, keeping the learner's attention on the material,
enforced rather than requested. A reader may reasonably object that assessment
gating and sequencing are ordinary course rules rather than monitoring, and
that grouping them with camera and microphone sensing inflates what monitoring
appears to cost. We keep the bundle because the learner meets it as a single
enforced experience and rarely separates the parts when writing about it, and
we report sensing and rule enforcement separately wherever a count turns on
the distinction (Section IV.C). Deliberately excluded from this scope are the
platform's general reliability issues, slot booking, content difficulty, and
administrative matters such as certificates, which surface in the same data
but concern different design questions (Section VI).

\subsection{Data Sources}

\begin{figure*}[t]
\centering
\begin{tikzpicture}[
  font=\small,
  node distance=6mm,
  box/.style={draw,rounded corners=1pt,align=center,inner sep=4pt,
              minimum height=8mm},
  stage/.style={box,fill=black!3,minimum width=38mm},
  form/.style={box,minimum width=58mm},
  chan/.style={box,minimum width=44mm,font=\footnotesize},
  lbl/.style={font=\footnotesize\itshape},
  arr/.style={-{Latex[length=2mm]},thick}
]
\node[stage] (onb) {\textbf{Onboarding course}};
\node[stage,right=16mm of onb] (ai) {\textbf{Fundamentals of AI}\\[1pt]
  \footnotesize first content course, approx.\ one week};
\node[stage,right=16mm of ai] (mern) {\textbf{MERN stack development}\\[1pt]
  \footnotesize persisting stage};
\draw[arr] (onb) -- (ai);
\draw[arr] (ai) -- (mern);

\node[form,below=12mm of ai] (fa) {\textbf{Form A $\cdot$ dropout exit survey}\\[1pt]
  \footnotesize 17 responses, 15 analyzable $\cdot$ 7 raise focus-monitoring};
\node[form,below=12mm of mern,xshift=6mm] (fb)
  {\textbf{Form B $\cdot$ persisting-learner survey}\\[1pt]
  \footnotesize 36 analyzable responses $\cdot$ 11 raise focus-monitoring};
\draw[arr] (ai) -- node[lbl,left=1mm] {did not continue} (fa);
\draw[arr] (mern) -- node[lbl,right=1mm] {progressed} (fb);

\node[chan,below=16mm of fa,xshift=-46mm] (tel)
  {\textbf{Camera-verification telemetry}\\ 14,529 events $\cdot$ 448 students\\
   median 20 flags per student};
\node[chan,right=6mm of tel] (emo)
  {\textbf{Emotion widget}\\ 615 events $\cdot$ 273 students\\
   131 carry free text $\cdot$ optional};
\node[chan,right=6mm of emo] (eoc)
  {\textbf{End-of-course survey}\\ up to 634 respondents per item\\
   6 Likert and 2 open items $\cdot$ mandatory};
\begin{scope}[on background layer]
  \node[draw,fill=black!3,rounded corners=1pt,fit=(tel)(emo)(eoc),
        inner sep=7pt,label={[font=\footnotesize\bfseries,inner sep=2pt]above:%
        Platform records, MERN course, read-only pseudonymized extract}]
        (plat) {};
\end{scope}

\node[box,below=10mm of emo,minimum width=60mm] (tri) {\textbf{Triangulation}\\[1pt]
  \footnotesize a finding counts only when it appears in at least two
  independently collected sources};
\draw[arr] (tel.south) |- ([yshift=-4mm]plat.south) -| (tri.north);
\draw[arr] (eoc.south) |- ([yshift=-4mm]plat.south) -| (tri.north);

\draw[arr] (fa.west) -- ([xshift=-5mm]plat.west |- fa)
                    -- ([xshift=-5mm]plat.west |- tri) -- (tri.west);
\draw[arr] (fb.east) -- ++(5mm,0) |- (tri.east);

\node[box,below=10mm of tri,xshift=-44mm,text width=62mm] (rq1)
  {\textbf{RQ1 $\cdot$ monitoring as support}\\[1pt]
  \footnotesize does it help learners stay focused and carry them to
  completion?};
\node[box,below=10mm of tri,xshift=44mm,text width=62mm] (rq2)
  {\textbf{RQ2 $\cdot$ monitoring as hindrance}\\[1pt]
  \footnotesize in what ways, and to what extent, does it stand in their
  way?};
\draw[arr] (tri.south) -- (rq1.north);
\draw[arr] (tri.south) -- (rq2.north);
\end{tikzpicture}
\caption{Study design. Two free-text instruments form the qualitative core,
and three platform channels collected independently of them test whether the
failure modes learners describe reproduce at scale. The free-text forms are not linked to the platform records, and the two
program stages are not pooled as one population. The three platform channels
are joined to one another per learner, through a stable pseudonym, for the
single correlation reported in Section IV.C; the platform's watch-time log is
joined the same way, since the flag rate is flags per minute watched.}
\label{fig:design}
\end{figure*}
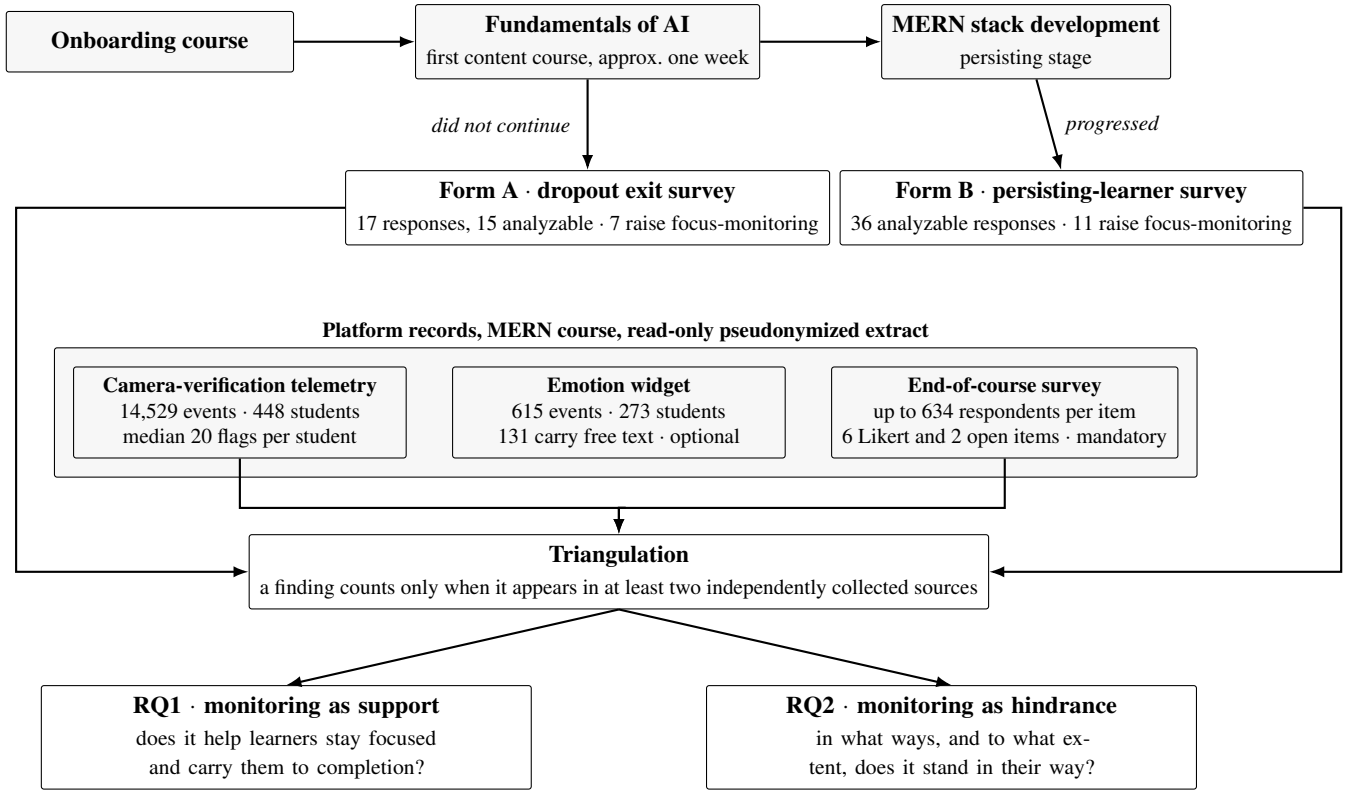

The research questions concern learners' subjective experience of
focus-monitoring and the extent to which that experience generalizes across
the cohort. Addressing both requires qualitative accounts from individual
learners alongside quantitative records collected by the platform over the
same program. Accordingly, five data sources inform this study, two
qualitative and three quantitative, summarized in Fig.~\ref{fig:design}.

The qualitative core is a pair of free-text instruments. Form A, a dropout
exit survey, was administered to learners who stopped engaging with the course
and received 17 responses, of which 15 were analyzable. Two were set aside:
one respondent reported submitting the form by accident, and one reported
having completed the course. Form B, a reflection survey, was administered to
learners who progressed beyond the first content course and yielded 36
analyzable responses. For both forms, responses submitted up to 20:05 IST (14:35 UTC) on
5 August 2026 were included. Both instruments are open-ended, and neither asked about
proctoring or monitoring by name. Every mention of focus-monitoring analyzed
in this study is therefore raised by the respondent rather than solicited by
the instrument. Table~\ref{tab:sources} lists each source with its type and the extent of the
data it contributed.

Three quantitative channels, drawn course-wide from the same program's MERN
course, the stage that follows the AI course, allow us to test whether what the
free-text respondents describe reproduces at scale. The camera-verification
telemetry logs 14,529 events across 448 students. The in-video emotion widget,
a five-point affect scale with optional free text that a learner can submit
for any content item, contributed 615 events from 273 students, 131 of them
carrying text. The mandatory end-of-course survey, six Likert items and two
open-text items, reached up to 634 respondents per item. All three channels were extracted on 4 August 2026 and cover the MERN
course from its first logged records to that date; the first camera flag was
logged on 24 May 2026 and the first emotion submission on 22 May 2026. These channels were collected independently of Forms A and B, under a
separate read-only, pseudonymized pipeline. Because the two pipelines
pseudonymize separately, we cannot establish how far their populations
overlap: Form B respondents had progressed into the MERN course and were
therefore eligible for the same end-of-course survey. We treat these channels
as independently collected instruments, not as disjoint populations.

\begin{table}[t]
\caption{Data Sources}
\label{tab:sources}
\centering\small
\begin{tabular}{L{0.30\columnwidth} L{0.20\columnwidth} L{0.36\columnwidth}}
\toprule
\textbf{Source} & \textbf{Type} & \textbf{n} \\
\midrule
Form A: dropout exit survey & Qualitative, free text & 15 analyzable
respondents \\[0.3em]
Form B: persisting-learner reflection survey & Qualitative, free text & 36
analyzable respondents \\[0.3em]
Camera-verification telemetry & Quantitative, event log & 14,529 events, 448
students \\[0.3em]
Emotion widget & Quantitative + free text & 615 events, 273 students \\[0.3em]
Mandatory end-of-course survey & Likert + free text & up to 634 respondents
per item \\
\bottomrule
\end{tabular}
\end{table}

\subsection{Analysis Procedure}

We follow the convergent mixed-methods logic of Creswell and Plano Clark
(2018): the qualitative and quantitative data are collected independently and
brought together at interpretation. Only
learners' own words can tell us how monitoring is experienced, and the paired dropout
and persisting-learner instruments reach the very population that
completer-only studies miss. Dropouts rarely answer surveys, so the
qualitative samples are small by nature, not by choice. We compensate through
methodological triangulation (Denzin, 1978): a failure mode counts as
established only when it reappears in independently collected channels.
Neither form named monitoring, so no mention was solicited by the wording of a
question. Both forms did ask respondents to describe what studying on the
platform was like, from opening the course to finishing a video, which leads a
respondent through the monitored steps of a session without naming them.
Unprompted, in this paper, therefore means not named by the instrument, not
free of every cue.

Each Form A response was classified by the mechanism it addresses and by the
role that mechanism played in the respondent's departure: the primary stated
cause, a secondary grievance beside another cause, or a low-confidence
mention. Each Form B response was classified by mechanism and by polarity:
complaint, praise, or explicitly mixed within the same answer. A response
entered the evidence base only if it named or clearly described one of the
mechanisms in our scope. Everything else, reliability bugs, content
difficulty, certificates, general motivation, was set aside however
substantial it was in its own right (Section VI). All data were thematically
analyzed by a single researcher, and inter-rater reliability was therefore not
assessed. We acknowledge this as a limitation, as independent analysis could
have provided additional evidence of consistency (Section
VI). The coding record, listing each coded response with its mechanism, role
or polarity, and supporting text, is available as supplementary material.

For the quantitative channels, we calculated the frequency and distribution of
flagged events. For the telemetry data, we aggregated counts of anomalies by
student to identify failure modes; for the emotion widget, we computed the
frequency and polarity of submissions per content item; and for the mandatory
survey, we analyzed the distribution of Likert scores to assess cohort-wide
reception.

Counts are reported with their corresponding denominators (e.g., 7 of 15) instead of standalone percentages. Since focus-monitoring was
intentionally addressed only in a subset of survey items, reporting 7 of 15 as
47\% could misleadingly imply that the finding represents the full analyzable
sample rather than a small, purposively scoped subset. Where we report a
Fisher exact test or a rank correlation, it serves as a descriptive check on a small sample and carries no power to test a hypothesis, so we draw no positive conclusion from a result that does not reach significance.

\subsection{Ethical Considerations}

Consent for monitoring is obtained at enrollment. Participation in the
platform required explicit user permission for camera and microphone access,
and a learner accepts a consent form and a proctoring declaration before the
first video plays. The course interface was restricted to a full-screen
environment, which also keeps learners continuously aware of the monitoring
state. For the purposes of this study, all platform records were extracted as
read-only, pseudonymized datasets, ensuring that no personally identifiable
information (PII) was accessible during analysis. The qualitative survey
responses were collected on a voluntary basis, and participants were informed
that their feedback would be used for research purposes while remaining
detached from their course grading or standing. The research data was stored
in a secure environment accessible only to the research team. Throughout the
thematic analysis, all participant identifiers were replaced with pseudonyms
(e.g., A01, B01) to ensure participant anonymity in the reported findings.
Research use of the platform records rests on the consent given at enrollment
rather than on a separate study-specific consent, and the study was not
reviewed by an institutional ethics board. We state both facts plainly and
return to them in Section VI.

\section{Findings}

\subsection{Who Spoke, and About What}

Of the 51 analyzable respondents, 18 raised focus-monitoring: 7 of 15 dropouts
and 11 of 36 persisting learners. A Fisher exact test finds no
significant difference between the two proportions ($p = 0.34$), which we read
descriptively rather than as a test of equivalence on samples this small. 

Eight of the 15 dropouts left for reasons unrelated to monitoring: removal
from the internship program, competing academic schedules, lack of time, or
technical failures such as videos that did not load. Focus-monitoring is
therefore a minority topic in both instruments, and it occupies both groups to
a similar degree. However, the two groups differ in what they say
about focus-monitoring.

\subsection{Monitoring as Support (RQ1)}

Table~\ref{tab:persisting} presents the 11 persisting-learner responses that address
focus-monitoring, with mechanism, polarity, and supporting text. Four
respondents write about the monitoring only positively, and two more praise it
within the same response that criticizes it. Each positive response names a
concrete function the mechanism serves: enforcing seriousness (B04), enabling
focus (B27), sustaining engagement (B35), and creating a working environment
before a session begins (B34). B36 states the self-regulation scaffold of
Section II in the learner's own words, keeping the video open ``meant i can
not multi task'', the platform providing the regulation the learner would not
provide alone.

The mandatory survey is consistent with this at cohort scale. The two items that
ask directly about the monitoring experience average above the scale midpoint,
3.64 for checkpoint-worth-it and 3.72 for ease-while-monitored, on a
five-point scale. A mean above the midpoint does not establish that a majority
endorse the mechanism; that claim needs the response
distributions. What the means do support is still
useful: at cohort scale the monitoring experience is not rejected outright,
while the free text shows why the same mechanism divides the learners who
write about it.

\subsection{Monitoring as Hindrance (RQ2)}

Of the seven dropouts who raise focus-monitoring, four name it as their
primary reason for leaving rather than a grievance beside another cause.
Table~\ref{tab:dropouts} lists the seven responses with the mechanism each
names and the role it played in the departure.

The four primary-cause departures divide into two categories. Two follow a
misreading of ordinary behavior. A05 was stopped at the environment check that
precedes the first video, flagged before engaging with any material, and did
not return. A03 describes a head movement sending an already-playing video
back to the start of its segment: ``if we just move of head we the video
playback get again started'' [sic]. The other two follow a rule working as
designed. A11 left over the score threshold that returns a learner to the
content sequence, and A12 over the restrictions on how a video may be played.
Neither was flagged in error; what the rule cost them was work they had
already finished. The split between camera sensing and rule enforcement
matters for what any design change could address.

The persisting learners name the same mechanisms. Five of the 11 in Table~\ref{tab:persisting} write about focus-monitoring only critically, and two more criticize it inside a response that also praises it. B08 reports the video restarting while drinking water. B32 reports an error on head movement followed by an automatic lock. B01 and B02 report flags raised when nothing had happened, a face recorded as absent or a second face recorded while the learner was alone. B21 accepts the detection and objects to what follows it, describing a system that notices a distraction and then ``plays everything from beginning''. Hence, both groups report the same triggering events, and they differ in what each event cost the learner.

\begin{figure}[t]
\centering
\includegraphics[width=\columnwidth]{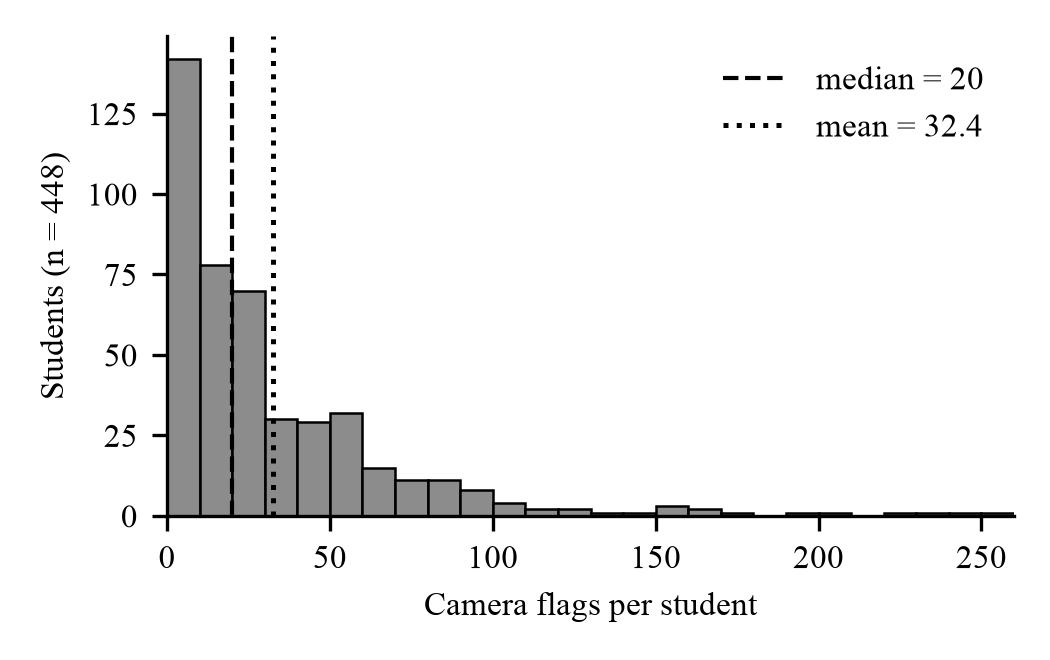}
\caption{Camera-verification flags per student ($n = 448$). The distribution
is right-skewed: most flagged learners fall below 30, while a tail extends to
253. Median 20, mean 32.4. Bins are 10 flags wide.}
\label{fig:flags}
\end{figure}

The camera telemetry shows how often the mechanism fires. The log holds
14,529 verification events from 448 students, and all but one are
face-recognition flags: the channel records the moments the system raised a
flag, never the moments it passed a learner through. Among the students who
appear in the log, the median student was flagged 20 times and the mean 32.4,
with the most flagged student reaching 253 (Fig.~\ref{fig:flags}). However, three
limitations apply to these figures. Learners who were never flagged do not appear in the log, so
these figures describe flagged students rather than the enrolled cohort; the
counts are not normalized by time spent watching; and the log records that a
flag was raised, never whether it was raised correctly, so it measures the rate
at which the mechanism fires, not the rate at which it errs. For a student
at the median, flags recur often enough to be a routine part of watching a
video.

The mandatory end-of-course survey reaches a larger population and its two
open items name the same triggers. Respondents describe a slight movement
being read as cheating, a water bottle misread while drinking and pausing the
lecture, and a warning not to speak appearing when there had been no sound.
One respondent describes the penalty score rising while alone and silent, and
the same lecture having to be watched again. These are the accounts of A03,
B08, and B32 arriving through a different instrument, though some Form B
respondents may also have answered it (Section III.C).

The emotion widget adds a third group, opted in item by item instead of being surveyed. Most of its 615 submissions from 273 students concern content and
platform reliability, outside the scope of this study. A small number fall
inside it, including a video that would not register as complete because of a
neighbor's noise, and an hour spent at quarter speed trying to reach the next
video. The microphone appears here and in the dropout accounts, and nowhere in
the persisting-learner responses.

How often a learner is flagged shows no detectable relationship with reported
discomfort. Across the 103 learners for whom a telemetry record and a survey response
could be brought together, the camera flag rate is uncorrelated with reported
ease while monitored ($\rho = -0.13$, $p = 0.18$) and with overall
satisfaction ($\rho = -0.003$). We report this as a null result and infer no mechanism from it. A sample of 103 gives little power to detect a small
association, and a cost driven by the severity of single events rather than by
their number would produce this pattern, but so would a sample too small to
reveal one.

\subsection{The Two Readings Together}

The two research questions produce complementary rather than competing
answers. A minority of persisting learners credit focus-monitoring with
providing the attention structure that a self-paced course otherwise leaves
entirely to the learner, while a majority of the dropouts who mention
monitoring name it as the primary reason they stopped.
Table~\ref{tab:convergence} brings these findings together across the five
data sources. The counts are small; what gives the convergence its value is
the independence of the sources. A failure mode that appears in a dropout's exit survey, in a
persisting learner's reflection, and again in a mandatory survey answered by a
different group is unlikely to be an artifact of one instrument or one
aggrieved respondent. Convergence of this kind establishes that the failure
modes recur across the program. It does not establish how often they occur,
which would need a denominator none of these channels provides.

\begin{table*}[t]
\caption{Focus-Monitoring Mentions Among Persisting Learners ($n = 11$ of 36)}
\label{tab:persisting}
\centering\small
\begin{tabular}{L{0.04\textwidth} L{0.14\textwidth} L{0.08\textwidth}
                L{0.62\textwidth}}
\toprule
\textbf{ID} & \textbf{Mechanism} & \textbf{Polarity} & \textbf{Evidence} \\
\midrule
B01 & Camera, face count & Complaint & ``my face is not visible or two faces
visible when I am alone'' \\
B02 & Camera, false flags & Complaint & ``it flags without any real
anomalies'' \\
B04 & Monitoring generally & Positive & ``like exam platform with high
security'' \\
B06 & Camera, sensitivity & Complaint & ``make camera tracking a bit more
lenient'' \\
B08 & Camera, restart & Complaint & ``even if i am drinking water also it is
restarting my course video back'' \\
B21 & Camera, restart & Mixed & ``camera detection is good but\ldots\ ever it
notice some distraction'' \\
B27 & Monitoring generally & Positive & ``felt very helpful to be focused and
learn things in a new way'' \\
B32 & Camera, motion and lock & Complaint & ``even if i move my head it\ldots\
it automatically gets locked'' \\
B34 & Environment check and face recognition & Mixed & ``creates a studious
environment and helps us focus''; ``While writing notes the face detection
raises identity issue'' \\
B35 & Monitoring generally & Positive & ``liked its proctoring system which
kept us engaged to do the course obediently'' \\
B36 & Monitoring generally & Positive & ``keeping the video open\ldots\ meant
i can not multi task'' \\
\bottomrule
\end{tabular}
\end{table*}

\begin{table}[t]
\caption{Focus-Monitoring in Dropout Responses\\($n = 7$ of 15)}
\label{tab:dropouts}
\centering\small
\begin{tabular}{L{0.07\columnwidth} L{0.46\columnwidth} L{0.32\columnwidth}}
\toprule
\textbf{ID} & \textbf{Mechanism} & \textbf{Role in departure} \\
\midrule
A03 & Camera, distraction-triggered restart & Primary stated cause \\[0.3em]
A05 & Environment check, flagged before content began & Primary stated cause
\\[0.3em]
A11 & Assessment gating & Primary stated cause \\[0.3em]
A12 & Sequencing, playback restriction & Primary stated cause \\[0.3em]
A01 & Microphone, false-positive penalty & Secondary grievance \\[0.3em]
A17 & Microphone, ambient noise & Secondary grievance \\[0.3em]
A09 & Sequencing, unspecified restrictions & Low-confidence mention \\
\bottomrule
\end{tabular}
\end{table}

\begin{table*}[t]
\caption{Convergence Across Sources}
\label{tab:convergence}
\centering\small
\begin{tabular}{L{0.22\textwidth} L{0.16\textwidth} L{0.26\textwidth}
                L{0.24\textwidth}}
\toprule
\textbf{Finding} & \textbf{Learner responses} & \textbf{Platform records} &
\textbf{Reading} \\
\midrule
Mid-video restarts are the signature friction & A03, B08, B21, B32 & Survey free text names the same triggers; telemetry records 14,529 flags,
with no restart field & Three self-reported sources; restarts are not logged \\[0.4em]
A flag raised before content begins can end participation & A05 & No channel
records the environment check & Learner accounts only \\[0.4em]
Microphone false positives penalize the home setting & A01, A17 & Survey free
text and emotion widget & Present in three sources, all self-reported \\[0.4em]
Cost is driven by severity rather than frequency & Objections cite single
events, not counts & Flag rate uncorrelated with reported ease ($\rho =
-0.13$, $p = 0.18$, $n = 103$) & Consistent with the accounts; the null result
does not establish it \\[0.4em]
Monitoring works as attention structure for a minority & B04, B27, B35, B36 &
Both monitoring items average above the scale midpoint & Accounts explain the
means; distributions not yet reported \\[0.4em]
No objection to monitoring in principle & None of the 18 responses & No such
objection in the free text examined & Stated as an observed absence \\
\bottomrule
\end{tabular}
\end{table*}

\section{Discussion}

The telemetry counts the
moments the system raised a flag, and no channel counts the sessions that
passed without one, so the ratio of uneventful to interrupted sessions is not
something this study can report; the platform does not record playback
sessions as a unit. However, the learner accounts show what a flag
interrupted and what it cost the learner. An ordinary
action, a learner reaching for a glass of water, is read as distraction, and
the video restarts.

The clearest result of this study is that such an episode never generalizes
into an objection to the monitoring system itself. Among the students who
raised focus-monitoring, none stated that the system should not exist. Several
stated the opposite, crediting it directly with helping them remain engaged.
This is an absence observed in a small, self-selected set of responses, and it
carries the weight an absence can carry: learners who reject monitoring
outright may never have enrolled, may have left before any survey reached
them, or may simply not have raised the subject in a free-text box. The
complaints that do appear converge on a narrower claim: certain ordinary
actions, drinking water among them, are misread as distraction, and the
response that follows is disproportionate to the misreading.

Following a restart, the learner must notice the interruption and work back to
the point already reached. At the flag frequencies reported in Section IV, the
cumulative cost of these corrections competes with the attention the system
was designed to preserve. This is where the guardrail becomes a barrier: the
mechanism built to protect attention now consumes it. We did not measure the
time a restart costs, so this
remains an inference from the accounts rather than a measured quantity.

A related effect is the erosion of trust in the system among learners who
continue to endorse its underlying purpose. Reliance
on an automated system tracks its demonstrated reliability, increasing when
the system performs as expected and declining when it does not (Lee \& See,
2004; Hoff \& Bashir, 2015). A learner need not be reasoning explicitly about
automation theory for this effect to hold. Each misclassification functions as
evidence, however small, that the system's judgment cannot be fully trusted,
and this evidence accumulates in step with the interruptions themselves. One
respondent's account illustrates both effects within a single statement,
describing the same system as one that ``creates a studious environment and
helps us focus'' and one that raises an identity flag while notes are being
written, an unambiguously on-task behavior. We collected no measure of trust,
and we offer this reading as an interpretation of the accounts; it is not a finding of the study.

Prior work has established that such systems produce measurable
anxiety and discomfort (Conijn et al., 2022; Woldeab \& Brothen, 2019), and
that students remain uneasy about what is observed and retained (Balash et
al., 2021; Marano et al., 2024). That literature has not distinguished between
two possible sources of this discomfort, the fact of being monitored, or the
specific experience of being monitored inaccurately. This distinction is what
the present study adds to the literature on remote monitoring. The accounts
here point to the latter, within the limits of who answered: the discomfort
documented in this study traces to misclassification and to the severity of
the response that follows it, rather than to surveillance as such.

Hence, the target of a design change is the response to a flag rather than
the monitoring itself. The strongest complaints in
this dataset, A03's restart after a head movement, B08's restart while
drinking water, B21's acceptance of the detection alongside an objection to
what follows it, and B34's endorsement of the environment the system creates
alongside a flag raised during note-taking, share a structure. The detection
event is treated as reasonable, or at least tolerable; the response to it is
not. The tipping point, on this evidence, is a proportionality boundary: a
response sized to the lapse holds the learner in the course, and a response
that takes back completed work pushes the learner out of it. The first lever
is therefore the severity of the response. Substituting a
dismissible pause for an irreversible restart would leave the learner in
control of the session while still marking the moment the system lost
confidence. The obvious objection is that an interruption a learner can
dismiss no longer enforces anything, and enforcement is precisely what the
learners in Section IV.B credit the platform with. A dismissible pause answers
that objection only if dismissals are themselves recorded and surfaced to the
instructor, so that the mechanism documents a suspected lapse instead of
punishing it, with the restart reserved for repeated or identity-level
anomalies. This is a design conjecture rather than a tested result. The study
identifies where the friction arises but does not test whether this change
would reduce it.

A second lever is detection tolerance at one specific point. B01, B02, B06 and
B32 describe flags raised when nothing had happened, so detection accuracy is
not beyond question, and A05 shows where an early false positive costs most: a
flag raised before any content has been seen reaches a learner with nothing
yet invested in the course, and that learner did not return.

Set against the broader literature, this study narrows a general
question: which design parameters, detection tolerance, response severity, and
the timing of false positives, account for the observed friction, and which
aspects already function as intended for the learners who value them. This
study can pose that question precisely, on evidence from learners who lived
with the mechanism for a course rather than for an examination. The prevalence of this friction across the
enrolled cohort remains unmeasured.

\section{Limitations}

The qualitative findings rely on a voluntary pilot sample ($n = 51$), and
because responses were unprompted, these data may not fully capture the
experiences of the wider cohort. As an exploratory study, these results are
descriptive; the limited sample size prevents inferential comparison between
dropouts and persisting learners, and the Fisher exact test and rank
correlations reported above are descriptive checks rather than powered tests.
Further, the study lacks a control group or pre-monitoring baseline, so
we cannot isolate the monitoring mechanism's independent effect on completion
from confounding factors such as course content, workload, or personal
circumstances.

Methodologically, the reliance on a single coder without inter-rater
reliability checks limits analytical independence. Triangulation is
constrained by data alignment: while the monitoring mechanisms were
consistent, qualitative and quantitative data are drawn from adjacent program
stages (AI vs.\ MERN), and platform telemetry by design describes only engaged
learners, under-representing those who disengaged earliest. The telemetry
records flags and not their correctness, so no quantitative channel in this
study can separate a false positive from a correct detection; the distinction
on which the argument turns rests on learner accounts alone. The
face-recognition matching threshold was tightened in a platform update merged
on 17 June 2026, and 14,511 of the 14,529 flags were logged after that date;
no flags were logged between 8 and 21 June. The flag counts therefore describe
the stricter setting. Finally, while
telemetry corroborates most failure modes, the absence of voice-event logging
during the data window means microphone-related claims rest solely on
qualitative accounts.

The study was not reviewed by an institutional ethics board, and research use
of platform records rests on the consent learners give at enrollment rather
than on a study-specific consent (Section III.E). Larger-scale, longitudinal
administrations with independent coding and prior ethical review are planned
for future cohorts.

\section{Conclusion}

This study examined how learners experience focus-monitoring when it operates
across an entire course rather than a single examination. Their objection is
to the accuracy of the mechanism and to the effects of its response, not to
monitoring itself. None of the 18 respondents who raised the mechanism
questioned being observed. Each objection concerned either an ordinary action
misread as distraction, or the size of the interruption that followed a flag.

RQ1 asked whether learners perceive focus-monitoring as a support layer, and a
minority do. Four of the 11 persisting learners who raised the mechanism
described it only positively, and each named a specific function it
performed, from preventing multitasking during playback to establishing a
working environment before a session begins. At cohort scale, both survey
items on the monitoring experience average above the scale midpoint, though
we cannot yet show that a majority holds this view.

RQ2 asked in what ways, and to what extent, focus-monitoring hinders learners.
Two failure modes account for the reported friction: in the first, an ordinary
movement is misread and the video returns to the start of the current
segment. In the second, a mechanism operates as designed and takes back work
the learner has already completed, as in assessment gating and sequencing
enforcement. Both modes appear among learners who left and among those who
continued, and the groups differ in the consequence each event carried.

On the overarching question of engagement and persistence, focus-monitoring
was the stated primary cause of departure for 4 of the 7 dropouts who raised
it. This makes it a reported cause of departure, not a measured determinant of
completion. Without a baseline or a comparison cohort, its effect cannot be
separated from the demands of the internship, course difficulty, and personal
circumstance.

Accordingly, the design recommendation is confined to how the platform
responds to a flag, and to one point of detection. What most needs revision is the response
rather than the sensor, with one exception: the environment check, where a
false positive reaches a learner who has not yet begun the material. An
interruption the learner can dismiss, a warning before a lockout, and more
tolerance at that check would address most of the friction documented here.

A classroom regulated attention because a teacher could observe a lapse and
respond in proportion. Focus-monitoring extends that function to the
self-paced online course. The evidence here indicates that its success depends
less on how accurately a system observes than on the proportionality of what
it does once it has observed. That proportionality marks the tipping point this
paper has moved towards without yet locating: the boundary at which an
attention guardrail becomes a barrier to learning.

Future work will scale the two-track instrument and map its responses against
the mandatory end-of-course survey. We will compare camera and voice logs
across learner groups, with independent verification, to test whether flag
frequency predicts departure, and investigate whether focus-monitoring itself
aids persistence and whether reported friction stems from learner
misunderstanding.

\section*{Acknowledgment}

The authors thank the VicharanaShala Lab for Education Design (VLED), Indian
Institute of Technology Ropar, for the opportunity to carry out this research
within its internship program, for access to the pseudonymized platform
records, and for supporting the study reported here. We also thank the
program's operations team and the learners whose responses made this study
possible.

\noindent\textit{Conflict of interest.} The authors developed ViBe, the
platform studied in this paper.

\noindent The authors used a generative AI assistant to help edit drafts of
the text and to format the manuscript. The authors designed the study, coded
the responses, checked all analyses, results, and references, and take full
responsibility for the content.

\section*{References}
{\small
\setlength{\parindent}{0pt}
\begin{list}{}{\leftmargin=1em \itemindent=-1em \itemsep=0.35em \parsep=0pt \topsep=0.4em}
\item Azhar, K. A., Iqbal, N., Shah, Z., \& Ahmed, H. (2024).
Understanding high dropout rates in MOOCs --- a qualitative case study from
Pakistan. \textit{Innovations in Education and Teaching International, 61}(4),
764--778.
\item Balash, D. G., Kim, D., Shaibekova, D., Fainchtein, R. A., Sherr,
M., \& Aviv, A. J. (2021). Examining the examiners: Students' privacy and
security perceptions of online proctoring services. \textit{Seventeenth
Symposium on Usable Privacy and Security (SOUPS 2021)}, 633--652.
\item Conijn, R., Kleingeld, A., Matzat, U., \& Snijders, C. (2022). The
fear of big brother: The potential negative side-effects of proctored exams.
\textit{Journal of Computer Assisted Learning, 38}(6), 1521--1534.
\item Creswell, J. W., \& Plano Clark, V. L. (2018). \textit{Designing
and conducting mixed methods research} (3rd ed.). SAGE Publications.
\item Deng, L., Zhou, Y., \& Broadbent, J. (2024). Distraction,
multitasking and self-regulation inside university classroom.
\textit{Education and Information Technologies, 29}(18), 23957--23979.
\item Denzin, N. K. (1978). \textit{The research act: A theoretical
introduction to sociological methods} (2nd ed.). McGraw-Hill.
\item Dewan, M. A. A., Murshed, M., \& Lin, F. (2019). Engagement
detection in online learning: A review. \textit{Smart Learning Environments,
6}, Article 1.
\item Hoff, K. A., \& Bashir, M. (2015). Trust in automation: Integrating
empirical evidence on factors that influence trust. \textit{Human Factors,
57}(3), 407--434.
\item Hutt, S., Krasich, K., Brockmole, J. R., \& D'Mello, S. K. (2021).
Breaking out of the lab: Mitigating mind wandering with gaze-based
attention-aware technology in classrooms. \textit{Proceedings of the 2021 CHI
Conference on Human Factors in Computing Systems}, 1--14.
\item Kizilcec, R. F., P\'erez-Sanagust\'in, M., \& Maldonado-Mahauad, J.
(2017). Self-regulated learning strategies predict learner behavior and goal
attainment in Massive Open Online Courses. \textit{Computers \& Education,
104}, 18--33.
\item Lee, J. D., \& See, K. A. (2004). Trust in automation: Designing
for appropriate reliance. \textit{Human Factors, 46}(1), 50--80.
\item Lepp, A., Barkley, J. E., Karpinski, A. C., \& Singh, S. (2019).
College students' multitasking behavior in online versus face-to-face courses.
\textit{SAGE Open, 9}(1).
\item Marano, E., Newton, P., Birch, Z., Croombs, M., Gilbert, C., \&
Draper, M. (2024). What is the student experience of remote proctoring? A
pragmatic scoping review. \textit{Higher Education Quarterly, 78}, 1031--1047.
\item Onah, D. F. O., Sinclair, J., \& Boyatt, R. (2014). Dropout rates
of massive open online courses: Behavioural patterns. \textit{Proceedings of
the 6th International Conference on Education and New Learning Technologies
(EDULEARN14)}, 5825--5834.
\item Pintrich, P. R. (1999). The role of motivation in promoting and
sustaining self-regulated learning. \textit{International Journal of
Educational Research, 31}(6), 459--470.
\item Reparaz, C., Azn\'arez-Sanado, M., \& Mendoza, G. (2020).
Self-regulation of learning and MOOC retention. \textit{Computers in Human
Behavior, 111}, 106423.
\item Robal, T., Zhao, Y., Lofi, C., \& Hauff, C. (2018a). Webcam-based
attention tracking in online learning: A feasibility study.
\textit{Proceedings of the 23rd International Conference on Intelligent User
Interfaces (IUI '18)}, 189--197.
\item Robal, T., Zhao, Y., Lofi, C., \& Hauff, C. (2018b). IntelliEye:
Enhancing MOOC learners' video watching experience with real-time attention
tracking. \textit{Proceedings of the 29th ACM Conference on Hypertext and
Social Media (HT '18)}, 106--114.
\item Wang, W., Zhao, Y., Wu, Y. J., \& Goh, M. (2023). Factors of
dropout from MOOCs: A bibliometric review. \textit{Library Hi Tech, 41}(2),
432--453.
\item Woldeab, D., \& Brothen, T. (2019). 21st century assessment: Online
proctoring, test anxiety, and student performance. \textit{International
Journal of E-Learning \& Distance Education, 34}(1), 1--10.
\item Wu, J.-Y. (2017). The indirect relationship of media multitasking
self-efficacy on learning performance within the personal learning
environment: Implications from the mechanism of perceived attention problems
and self-regulation strategies. \textit{Computers \& Education, 106}, 56--72.
\item Zimmerman, B. J. (2000). Attaining self-regulation: A social
cognitive perspective. In M. Boekaerts, P. R. Pintrich, \& M. Zeidner (Eds.),
\textit{Handbook of self-regulation} (pp.\ 13--39). Academic Press.
\end{list}}

\end{document}